\documentclass[twocolumn]{article}

\usepackage{graphicx}
\usepackage{float}
\usepackage{amssymb,amsmath}
\usepackage{wrapfig}
\usepackage{rotating}
\usepackage{array}
\usepackage{url}
\usepackage[numbers,sort&compress]{natbib}
\usepackage[scale=0.83]{geometry}

\usepackage{titlesec}

\def \aj {Astron. J.}
\def \mnras {Mon. Not. R. Astron. Soc.}
\def \apj {Astrophys. J.}
\def \apjs {Astrophys. J. Suppl. S}
\def \apjl {Astrophys. J. Lett.}
\def \aap {Astron. Astrophys.}

\def \pasj {PASJ}

\titlespacing\section{0pt}{12pt plus 4pt minus 2pt}{0pt plus 2pt minus 2pt}
\titlespacing\subsection{0pt}{12pt plus 4pt minus 2pt}{0pt plus 2pt minus 2pt}
\titlespacing\subsubsection{0pt}{12pt plus 4pt minus 2pt}{0pt plus 2pt minus 2pt}

\title{\textbf{The diverse evolutionary pathways of post-starburst galaxies}}
\author{M. M. Pawlik$^{1}$, S. McAlpine$^{2,3}$, J. W. Trayford$^{3,4}$, V. Wild$^{1}$, R. Bower$^{3}$,\\ R. A. Crain$^{5}$, M. Schaller$^{4}$, J. Schaye$^{4}$}
\date{
\footnotesize{
\vspace{-5pt}
$^{1}$School of Physics and Astronomy, University of St Andrews, North Haugh, St Andrews, KY16 9SS, U.K. (SUPA)\\
$^{2}$Department of Physics, University of Helsinki, Helsinki, Finland\\
$^{3}$Institute for Computational Cosmology, Department of Physics, Durham University, South Road, Durham, DH1 3LE, UK\\
$^{4}$Leiden Observatory, Leiden University, P.O. Box 9513, 2300 RA Leiden, the Netherlands\\
$^{5}$Astrophysics Research Institute, Liverpool John Moores University, 146 Brownlow Hill, Liverpool, L3 5RF, UK\\}
\vspace{5pt}
\normalsize{May 2018}}

\begin{document}
%\twocolumn[
%  \begin{@twocolumnfalse}
    \maketitle
%    \vspace*{-1cm}
%\begin{abstract}
%  Guideline for main article says no references

{\textbf{About 35 years ago a class of galaxies with unusual spectral characteristics was discovered in distant galaxy clusters \citep{DresslerGunn1983,CouchSharples1987}. These objects, alternatively referred to as post-starburst, E+A or k+a galaxies, are now known to occur in all environments and at all redshifts \citep{Zabludoff+1996,Yan+2009,Wild+2016,Paccagnella2017,Socolovsky+2018}, with many exhibiting compact morphologies and low-surface brightness features indicative of past galaxy mergers \citep{Zabludoff+1996, Almaini+2017}. 
They are commonly thought to represent galaxies that are transitioning from blue to red sequence, making them critical to our understanding of the origins of galaxy bimodality \citep{Tran+2003,Yang+2008,Wild+2009,Swinbank+2012,Wong+2012,Pawlik+2016}. However, recent observational studies have questioned this simple interpretation \citep{Dressler+2013, Pawlik+2018}. From observations alone, it is challenging to disentangle the different mechanisms that may lead to the post-starburst phase. Here we present examples of four different evolutionary pathways through the post-starburst phase found in the EAGLE cosmological simulation \citep{Schaye+2015,Crain+2015}: blue$\rightarrow$blue, blue$\rightarrow$red, red$\rightarrow$red and truncated. Each pathway is consistent with scenarios hypothesised for observational samples \citep{CouchSharples1987,Dressler+2013,Abramson+2013,Pawlik+2018}. The fact that post-starburst spectral characteristics can be attained via various evolutionary channels explains the diversity of properties of post-starburst galaxies found in observational studies, and lends support to the idea that slower quenching channels are dominant at low-redshift \citep{Trayford+2016,Rowlands+2018}}}

Most galaxies in the local Universe are either blue, gas-rich systems featuring a prominent disky component, where new stars are being formed from the interstellar gas reservoirs, or red, often spheroidal galaxies with little to no cold gas. The build up of the red population over cosmic time indicates that star-forming spirals are transformed into quiescent ellipticals.
The paucity of intermediate ``green valley" galaxies may imply that galaxies preferentially undergo a rapid blue-to-red transformation, rather than gradually consuming their gas supply via steady star formation. Numerous scenarios have been proposed that could lead to such a rapid transformation of galaxies, with or without concurrent alteration of morphology from disk to spheroid;
however, observations have found it difficult to corroborate or falsify these theories.  

Observations of post-starburst galaxies opened a new door into the study of galaxy evolution as their characteristics place them in a transition phase between the two main classes of the general galaxy population \citep{Tran+2003,Yang+2008,Wild+2009,Swinbank+2012,Wong+2012,Pawlik+2016}. 
Idealised simulations of galaxy mergers confirm that a brief ($\lesssim600$\,Myr) post-starburst phase can follow a gas-rich merger, in which a powerful central starburst is succeeded by rapid quenching of star formation and significant morphological transformation \citep{Bekki2005,Wild+2009,Snyder+2011}. However, a large fraction of local post-starburst galaxies contain considerable cold gas reservoirs that may
return them to the blue sequence \citep{Rowlands+2015,French+2015}, although the gas may not be in a favourable state for star formation \citep{French+2018a}. Additionally, those in the densest environments may reflect stages in a cyclic evolution within the red population \citep{Dressler+2013,Abramson+2013, Pawlik+2018}.
growth of the red sequence \citep{Wild+2009,Wild+2016,Rowlands+2018}.

The scarcity of post-starburst galaxies, particularly in the local Universe, makes disentangling their origins challenging. While the spectral features remain visible for up to $\approx1$Gyr, any tidal features from past mergers fade rapidly \citep[within 200-400\,Myr for SDSS-depth images][]{Pawlik+2016,Pawlik+2018}. However, with cosmological simulations now showing good agreement with a large range of galaxy population observables \citep[e.g.][]{Schaye+2015,Pillepich+2018}
watching post-starburst galaxies evolve in real-time has become an attainable goal.

In the Methods section below we describe how we identified galaxies  with post-starburst spectral signatures and M$_{\star}>10^{9.5}$M$\odot$  from the EAGLE simulation. We use a Principal Component Analysis (PCA) method that is sensitive to older and weaker bursts than traditional methods, and includes galaxies that do not completely shut-down their star formation following the burst. At $z=0.1$ we find a number density in EAGLE of $7\times10^{-6}\,{\rm Mpc}^{-3}$, consistent with an observational comparison sample. 

We followed the spectral evolution of all the $z=0.1$ snapshot galaxies over the previous 2.5\,Gyr, and identified 124 galaxies that exhibited post-starburst signatures during this time period. Of these galaxies, we identified three distinct evolutionary pathways consistent with those suggested by observational studies \citep{Dressler+2013,Pawlik+2018}. We found $\approx$27\% of the post-starburst galaxies transferred from blue to red sequence, 60\% remained in the blue cloud  typically with decreased specific star formation rates (sSFR), and 2\% were in a cyclic evolution within the red sequence. The remaining $\approx$10\% either start or end in the post-starburst phase. We additionally found that $\approx$50\% of galaxies entering the post-starburst region had not undergone a distinct short-lived starburst, but rather a sharp truncation of their star formation \citep{CouchSharples1987}. These truncated star-formation histories have been identified as potential progenitors of red-spiral and S0 galaxies \citep{Balogh1999}. 

Here we present the detailed evolutionary histories of three post-starburst galaxies, selected to represent the three different pathways, plus an example of a galaxy that has undergone truncation of its star formation. In Figure \ref{fig:evolution} we show their star-formation and merger histories, the evolution of their gas to stellar mass ratios, their shape-asymmetry parameter $A_{s}$ and concentration index $C$. The shape-asymmetry measures morphological disturbance with $A_s>0.2$ identifying past or ongoing mergers, while the concentration index takes values of around 3 and 4 for disks and spheroids respectively.  In Figure \ref{fig:images} we show synthetic $gri$ composite images of the four galaxies. Figure \ref{fig:PCA_evolve} presents the evolution of the two spectral indices used to identify the post-starburst galaxies: PC1 measures the shape of the continuum around $4000$\AA\,(a proxy for mean stellar age), and PC2 measures the excess Balmer line absorption. The excess of A and F stars in post-starburst galaxies leads to large PC2 values a few hundred Myr following the starburst. Further details are provided in the Methods section below.

\emph{\bf Case I - ID17473042 - ``Classic'' blue$\rightarrow$red.}
Prior to entering the post-starburst phase, object 17473042 is a star-forming galaxy with a star-formation rate (SFR) of $\approx5\mbox{M}_{\odot}$/yr and sSFR of $3\times10^{-10}$/yr.  It experiences a strong starburst about 1\,Gyr before $z=0.1$ associated with a major merger with another galaxy (stellar mass ratio of 0.62). The brief enhancement in star formation is followed by a rapid quenching leading to a quiescent remnant with sSFR$\approx10^{-12}$/yr by $z=0.1$. This decline in SFR is mirrored in the evolution of the cold gas mass to stellar mass ratio, which drops below 0.05 from values of $\approx$0.3 prior to the merger-induced starburst. In the PC1-PC2 distribution the galaxy undergoes a clear transition: starting out in the blue-cloud, then moving to smaller PC1 (higher sSFR) upon the onset of the starburst, passing through the post-starburst phase with high PC2, to rapidly head towards the red sequence. The transition is also visible in the morphology of the system: the violent interaction disrupts the stellar orbits from the initial disk-like configuration ($C<3.0$) leading to a brief but significant morphological disturbance ($A_{s}>0.2$). The total time of visible disturbance is about 0.5\,Gyr predominantly prior to the merger, after which the remnant settles into a system with a prominent spheroidal component ($C\approx4.0$). The image analysis reveals a second period of morphological disturbance at a lookback time of 3\,Gyr, with no corresponding interaction. However, a movie reveals a close rapid encounter (``fly-by"). 

\emph{\bf Case II - ID9052812 - blue cycle.}
Similar to the previous example, object 9052812 is a star-forming galaxy with a SFR of $\approx4\mbox{M}_{\odot}$/yr and sSFR of $\approx10^{-9}$/yr prior to entering the post-starburst phase. It experiences a strong starburst about 2\,Gyr before $z=0.1$ which coincides with a merger with a mass ratio of 0.26. Unlike the previous case the remnant continues to form stars after the starburst, with a reduced sSFR$\approx4\times10^{-11}$/yr by $z=0.1$. A large fraction of the galaxy's total gas reservoir is either exhausted or expelled during the merger; however, the residual gas is sufficient to continue star formation after the interaction. 
The evolution of the system in PC1-PC2 space is cyclic within the blue cloud (`blue cycle'), with a brief transition through the starburst and post-starburst regions. Similarly to before, the merger signatures are clearly visible for a brief period of time ($\approx0.5$\,Gyr) and growth of the spheroidal component occurs as the stellar orbits settle after the merger (from $C\approx3.0$ to $C\approx 4.0$). This case also clearly illustrates how a central starburst can temporarily increase $C$ to values $>4$.

\emph{\bf Case III - ID19955965 - red cycle.}
Object 19955965 (M$_{\star}$/M$_{\odot}$=10$^{11.3}$) is an example of a red-sequence galaxy with a sSFR of $3\times10^{-12}$ that experiences an enhancement in SFR following a minor merger (mass ratio of 0.12) with a gas-rich satellite. This results in an increase in the  gas mass fraction and corresponding increase in star formation activity for $\approx1.5$\,Gyr, after which the star formation is rapidly quenched.  This evolution within the red sequence is apparent in the PC1-PC2 space, where an originally quiescent galaxy moves to the star-forming region, after which it briefly passes through the post-starburst phase before returning  to the red sequence. The image analysis shows an already high central concentration of light in the galaxy prior to the merger, $C\approx4.0$. Neither a significant growth of the spherical component nor an obvious morphological disturbance is observed during the interaction. 

\emph{\bf Case IV - ID8154304 - Truncation.}
The final case is an example of a post-starburst phase which does not follow an evident starburst. These typically only just attain sufficiently strong Balmer lines to appear above the main sequence in PC2, similar to the ``red cycle" objects. Object 8154304 is a star-forming galaxy with SFR prior to the quenching event of $\approx1\mbox{M}_{\odot}$/year. It experiences rapid quenching about 0.5\,Gyr before $z=0.1$ associated with fast exhaustion of the galaxy gas supply. Three micro-mergers (mass ratio $<1:10$) occur in close succession, within 0.6\,Gyr of the quenching event. Additionally, the sudden increase in the the halo mass of the galaxy, from M$_{\mbox{h}}$/M$_{\odot}\approx10^{11.6}$ to M$_{\mbox{h}}$/M$_{\odot}\approx10^{14.4}$ at $z=0.1$, indicates that it has fallen into a massive halo. Analysis of the galaxy's environment also revealed a rapid decrease in its halo-centric distance. This suggests that the galaxy has been subjected to cluster-related quenching, including ram-pressure stripping of gas and harassment. In PC1-PC2 space, the galaxy undergoes a transition from the blue cloud towards the red sequence with a very brief period in the post-starburst phase. The quenching is not associated with growth of a spheroidal component and the galaxy retains a disk-like central concentration of light ($C\lesssim3.0$). Interestingly, the combined effect of multiple micro mergers and/or infall onto the cluster induces more morphological disturbance than the single minor merger observed in Pathway III.

\vspace{0.2cm}
Simulated galaxies in the $z=0.1$ EAGLE universe follow various pathways through the post-starburst phase closely resembling those conjectured from observational studies. A gas-rich major merger of two galaxies can cause a powerful starburst and subsequent decline in SFR (ID17473042 and ID9052812). This scenario agrees with the disturbed morphologies found in many low-redshift post-starburst galaxies \citep[e.g.][]{Zabludoff+1996,Yang+2008,Pawlik+2016}, 
as well as their intermediate-density environments which are favourable for mergers to occur \citep[e.g.][]{Zabludoff+1996, Yan+2009,Pawlik+2018}. Moreover, the short timescales of morphological disturbance post-merger supports the possibility of a merger origin for post-starburst galaxies with no visible signs of interaction \citep{Pawlik+2016}. Evolution via major mergers can lead the galaxy either back to the blue cloud with decreased sSFR (ID9052812) or near/onto the red sequence (ID17473042), depending on the remaining gas content \citep[e.g.][]{French+2015,Rowlands+2015,French+2018}.
detectable as a post-starburst galaxy, conforming to the findings of significant gas reservoirs in some post-starburst galaxies \citep[e.g.][]{French+2015,Rowlands+2015}. 
In both cases, the merger leads to growth of the spheroidal component, as found in idealised merger simulations \citep[e.g.][]{Lotz+2008}, 
and in agreement with many morphological studies of post-starburst galaxies \citep[e.g.][]{Mendel+2013,Almaini+2017,Pawlik+2018}.
As such, the quenching (whether partial or `full') and morphological transformation occur concurrently, as a result of the dynamical interaction between the galaxies. The starbursts are accompanied by a brief period of black hole growth, however, it is not currently possible to tell whether this contributes to the star-formation quenching.  

Star formation in a galaxy may also be quenched, either fully or partially, \emph{without} the corresponding morphological transformation (ID8154304). In this particular case, the rapid decline in the SFR is due to environmental effects, consistent with that proposed in clusters \citep[e.g.][]{CouchSharples1987,Dressler+1999, Balogh1999, Paccagnella2017,Socolovsky+2018}.  
This process naturally explains why morphological studies of post-starburst galaxies in clusters have reported a high incidence of disks \citep[e.g.][]{Oemler+1997}.

Finally, post-starburst characteristics can also be observed when an already quiescent galaxy undergoes a minor merger with a gas-rich satellite (ID19955965). This `rejuvenation' event rapidly consumes the new gas supply only to return to the quiescent state, consistent with observational results in the field \citep{Pawlik+2018} and clusters \citep{Dressler+2013}.

In this work we find that 60\% of galaxies with post-starburst features remain in the blue cloud within the 2.5\,Gyr time window that we probe. About 50\% of galaxies had not undergone a distinct short-lived starburst, but rather a sharp truncation of their star formation. These multiple pathways explain the wide variety in the observed properties of post-starburst galaxies. They may also in part explain the disagreement regarding the role of this peculiar class of objects in the build up of the red sequence: if a significant fraction of post-starburst galaxies are not in fact transitioning from the blue cloud to red sequence as previously believed, more room will be opened for slower quenching routes \citep{Wild+2009, Trayford+2016, Wild+2016, Rowlands+2018}. Future work will investigate which factors are dominant in determining these pathways (e.g. gas fractions, progenitor morphology, orbital configuration, environment, black hole feedback), as well as whether cosmological simulations capture the correct balance of evolutionary pathways.

\vspace{0.5cm}
Correspondence and requests for materials should be addressed to Dr Vivienne Wild (vw8@st-andrews.ac.uk). 

\section*{Acknowledgements}
MMP and VW acknowledge support of the European Research Council (SEDMorph, PI Wild). RAC is a Royal Society University Research Fellow. VW would like to thank Ariel Werle for his help investigating the PC amplitude offsets.  This work was supported by the Science and Technology Facilities Council (grant number ST/P000541/1). This work used the DiRAC Data Centric system at Durham University, 
operated by the Institute for Computational Cosmology on behalf of the 
STFC DiRAC HPC Facility (\url{www.dirac.ac.uk}). This equipment was funded 
by BIS National E-infrastructure capital grant ST/K00042X/1, STFC capital 
grant ST/H008519/1, and STFC DiRAC Operations grant ST/K003267/1 and 
Durham University. DiRAC is part of the National E-Infrastructure.

\section*{Statement of contributions}
MMP led the analysis and writing of the manuscript, with significant input from SM and JWT who provided the EAGLE data and associated analysis products. VW conceived the initial idea, supported MMP throughout the project and led the response to the referees reports. RB, RAC, MS and JS are builders of the EAGLE simulation. They read and commented on the manuscript.

\begin{figure*}
\centering
    \includegraphics[scale=0.8]{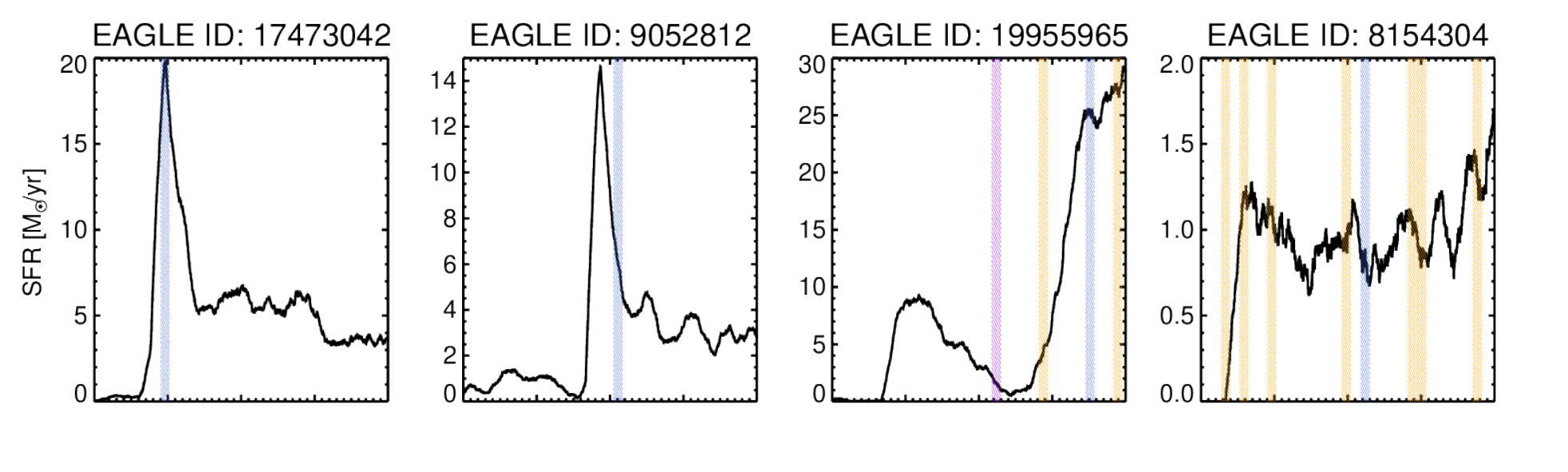}
    
    \vspace*{-1cm}
    
    \includegraphics[scale=0.8]{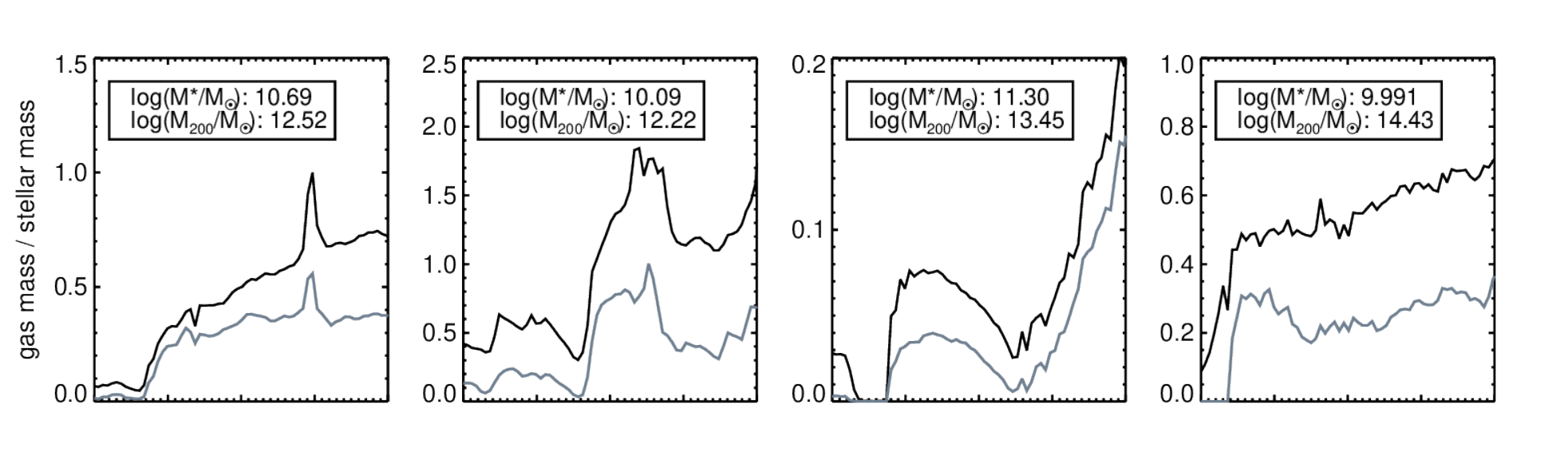}
    
    \vspace*{-1cm}
    
    \includegraphics[scale=0.8]{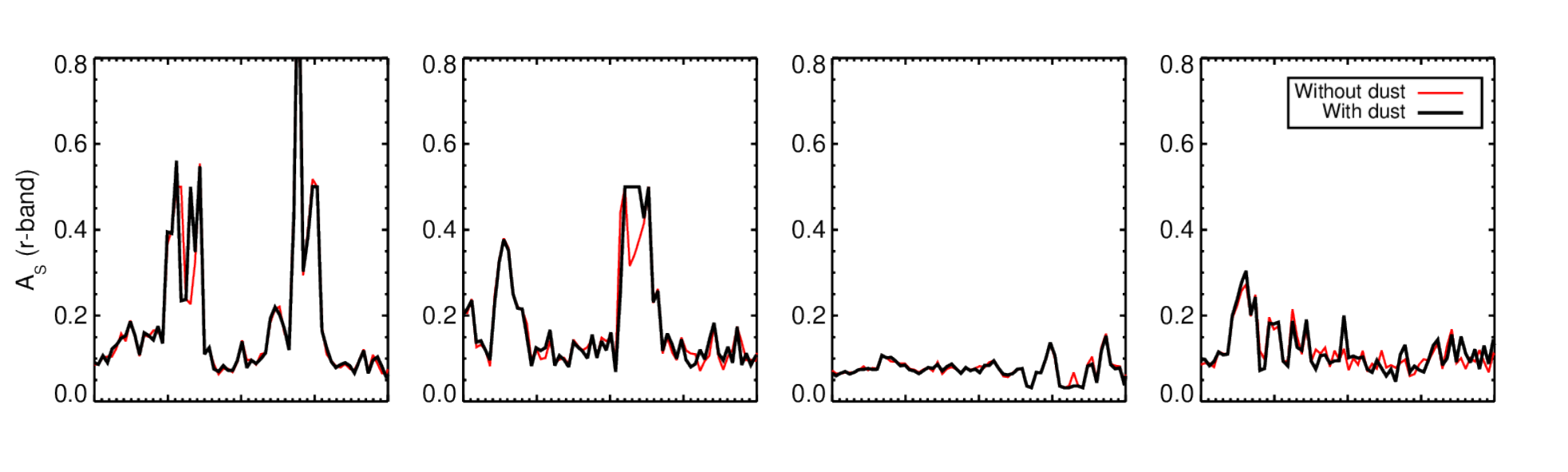}
    
    \vspace*{-1cm}
    
    \includegraphics[scale=0.8]{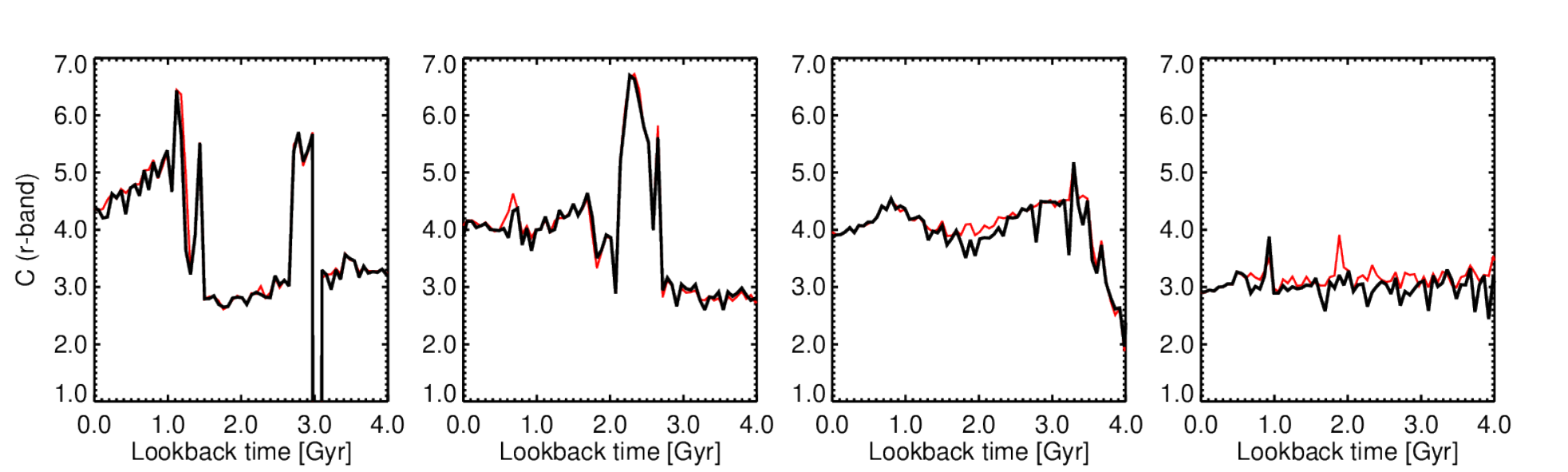}

\caption{The evolution of the star-formation rate, gas content and morphology of the four example EAGLE post-starburst galaxies as a function of lookback time from $z=0.1$. \emph{Top:} star-formation rate; the times at which mergers occur are marked with different colours according to the mass ratios of the progenitor galaxies: major mergers (1:1 - 1:4) in blue, minor mergers (1:4 - 1:10) in purple, and micro mergers ($<$1:10) in yellow. \emph{Upper centre:} the cold (gray) and total (black) gas to stellar mass ratio histories. The legend gives the stellar and dark matter halo mass for each galaxy at $z=0.1$. Stellar masses are calculated within 30\,kpc, and the halo masses are with respect to critical density. \emph{Lower centre and bottom:} the shape asymmetry and  concentration index measured using SDSS-like $r$-band images. Measurements made using dust-free and dusty images produced by SKIRT are shown in red and black respectively.
}
\label{fig:evolution}
\end{figure*}

\begin{figure*}
\centering
%\hspace*{17pt}
\includegraphics[scale=0.78]{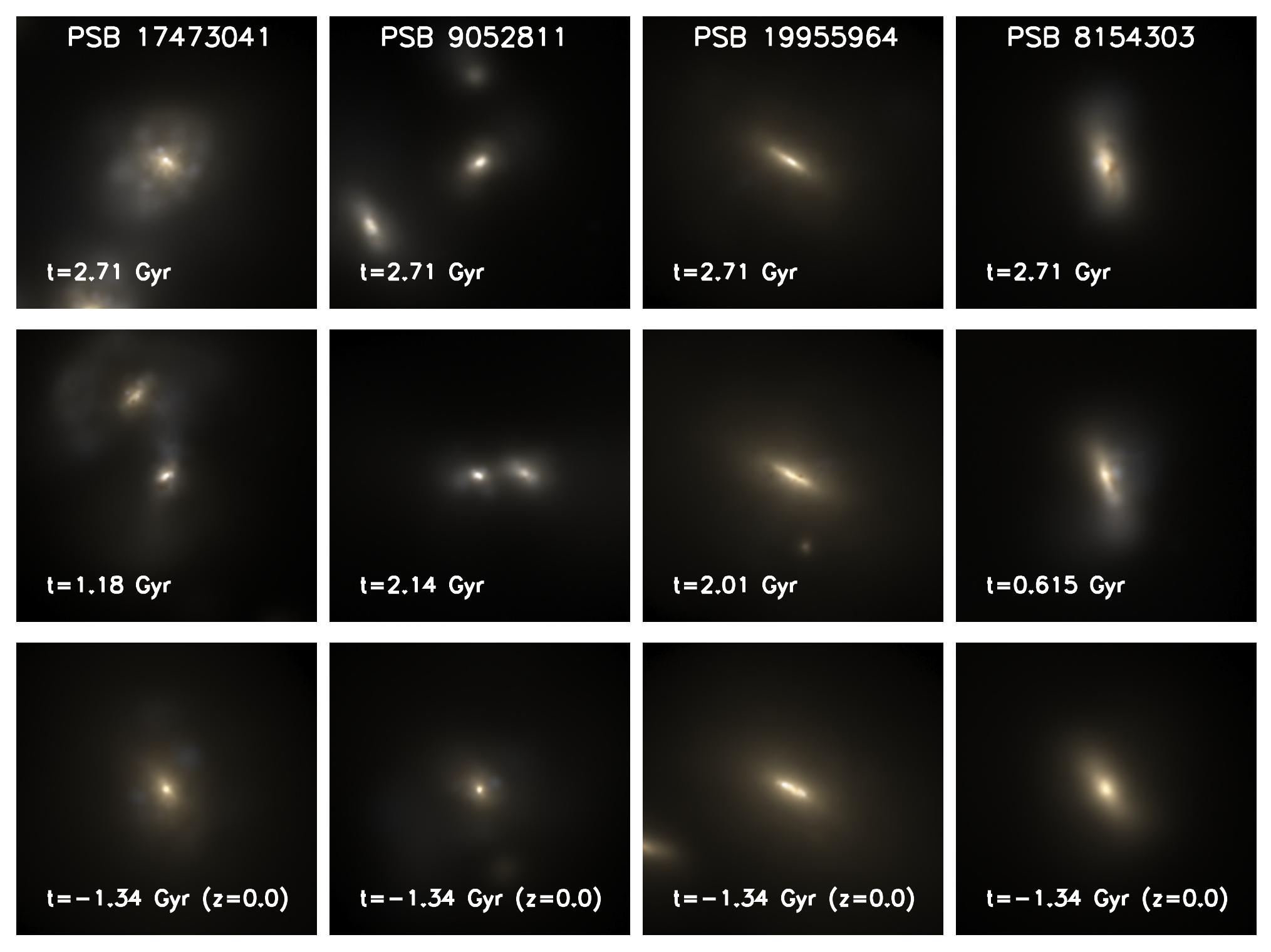}
\caption{SDSS-like (\emph{gri}) synthetic images of the EAGLE post-starburst galaxies showing their morphology at three different points in time: prior, during and after the merger-induced starburst (or quenching event in the case of 8154303). The quoted values of $t$ are lookback time from $z=0.1$. The images were generated using the SKIRT radiative transfer code, and include attenuation by dust. The field of view is 82\,kpc.}
\label{fig:images}
\end{figure*}

\begin{figure*}
\centering
 \includegraphics[scale=0.85]{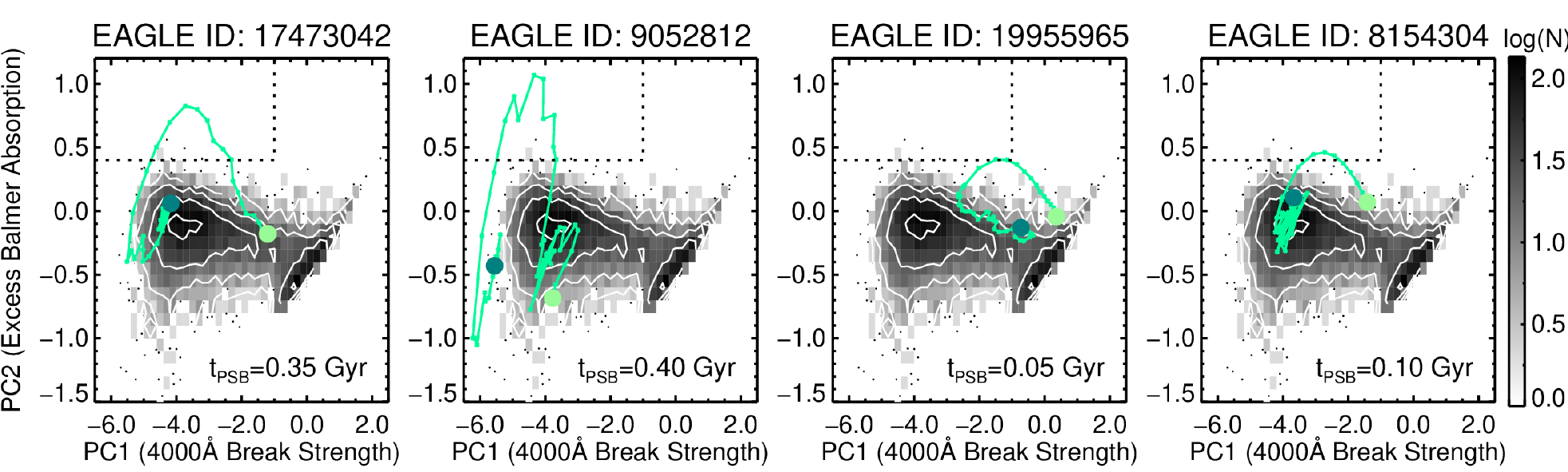}
 \caption{The evolution of the spectral indices, PC1 and PC2, for our four example galaxies. The value measured at $z=0.1$ ($t_{lb}=0.0$) is indicated by a light green circle; measurements are recorded in steps of 50\,Myr and trace the evolution of PC1 and PC2 back to $t_{lb}=2.5$\,Gyr (dark green circle).  The underlying greyscale shows the full EAGLE sample at $z=0.1$. The time that the galaxy spends in the PSB selection box is given in the lower right of each panel. }
 \label{fig:PCA_evolve}
\vspace{-10pt}
\end{figure*}

\newpage

\section*{Methods}

\subsubsection*{The EAGLE simulation and data products}

The EAGLE cosmological simulations are described in detail in  Schaye et al. \citep{Schaye+2015}; here we highlight some key points. The suite includes a range of cubic volumes and resolutions performed with the Gadget-3 TreeSPH code \citep{Springel2005}. The subgrid physics used in EAGLE includes radiative cooling, star formation, black-hole gas accretion, and AGN feedback \citep{Schaye+2015,Rosas-Guevara2015}, based on that used in OWLS \citep{Schaye+2010}, GIMIC \citep{Crain+2009} and COSMO-OWLS \citep{LeBrun+2014}.
The simulations were calibrated to match the observed galaxy sizes and galaxy stellar mass function (GSMF) at $z\approx0.1$ \citep{Crain+2015}. 
As in this work we are interested in local post-starburst galaxies, which are relatively rare, we focus on the largest EAGLE volume, Ref-L100N1504, with a comoving volume of $10^6$\,Mpc$^{3}$, contains $2\times1504^{3}$ particles and assumes a $\Lambda$CDM cosmology with parameters derived from the initial Planck data release, i.e. with $\Omega_{b}=0.0482$, $\Omega_{dm}=0.2588$, $\Omega_{\Lambda}=0.693$, and H$_{0}= 67.77$\,km\,s$^{-1}$\,Mpc$^{-1}$ \citep{Planck+2014}.

The EAGLE data products \citep{McAlpine+2016} used in this paper include the dark matter, stellar and gas mass histories as well as the merger trees. Full particle  information is stored for 29 snapshots over the redshift range $0\leq z \leq 20$,  spaced between $\approx$0.3 to $\approx$1.35\,Gyr from high to low redshift. In between these snapshots are finely spaced ``snipshots'' for which less information is stored. Galaxies are defined as all material within a 30\,kpc (proper) spherical aperture centred on the subhalo potential minima in the $z=0.1$ snapshot. As subhalo information is not stored for snipshots, we track the galaxy centre by tracing the most bound star particles from snipshot to snipshot and locating the centre using a shrinking spheres algorithm. Our sample is selected from galaxies with M$_{\star}>10^{9.5}$M$_\odot$ in the $z=0.1$ snapshot. This minimum mass limit corresponds to the point where the colour distribution of EAGLE galaxies diverges from observations \citep{Furlong+2015}, as well as corresponding to mass limits typically used in observations of local post-starburst galaxies \citep{Pawlik+2018}. We identify mergers for galaxies above a mass limit of 100 particles (M$_{\star}>10^{8}$M$_\odot$).

We computed the star-formation histories by summing up the stellar mass formed in each of the 12470 1\,Myr-wide bins of lookback time for $z>0.1$ (12.47\,Gyr of cosmic history), and the metallicity histories by finding the ratio of stellar mass locked up in metals to the total mass in stars for each age bin. To compute the histories of the total and cold gas we consider gas particles that are bound to the halo and within a 30\,kpc (proper) aperture of the galaxy. The cold gas component is defined to be those gas particles that are forming stars.  We choose to identify post-starburst galaxies in the $z=0.1$ snapshot to allow us to observe the evolution of the galaxies over a longer time period following the starburst/quenching episode than would be possible for those identified in the $z=0$ snapshot. We therefore define lookback time $t_{lb}=0.0$ from the $z=0.1$ snapshot.

\subsubsection*{Spectral synthesis and analysis}

We created optical synthetic SEDs for the 7312 EAGLE galaxies with stellar masses M$_{\star}>10^{9.5}$M$_\odot$ in the $z=0.1$ snapshot by matching the star-formation and metallicity histories with BC03 simple stellar populations \citep[SSPs,][]{BruzualCharlot2003}, interpolating the SSPs linearly in log-flux between metallicity bins. We used an updated (2016) version of the BC03 SSPs with a Chabrier Initial Mass Function (IMF) constructed using Padova 1994 stellar evolutionary tracks, and the MILES spectral library in the wavelength range of relevance to this paper \citep{miles2006}. 
We then applied a two-component dust screen assuming an optical depth to young stellar populations (ages $<10^7$ years) drawn randomly from a Gaussian distribution of width 0.5 centred on $\tau_V=1.0$, which approximately matches the dust distribution measured from Balmer line decrements for galaxies in our SDSS comparison sample (see below). We assume a fraction of dust due to the interstellar medium of $\mu=0.3$, and power-law slopes of 0.7 and 1.3 for the interstellar and birth cloud dust, respectively \citep{daCuhna+2008,Wild+2007}. The two-component model means that galaxies with no young stellar populations will have an effective attenuation of $\tau_V\approx0.3$. The more sophisticated radiative-transfer treatment used to construct the images (see below) was computationally too expensive for the construction of the spectra, but previous results on the EAGLE simulation suggest that an analytic two component dust prescription leads to a similar colour distribution compared to the full radiative transfer approach \citep{Trayford+2017}. 

In observational studies post-starburst galaxies are selected based on their unique spectral energy distribution (SED) characteristics. i.e. strong Balmer absorption lines or Balmer break \citep{DresslerGunn1983, Goto+2003, Wild+2007, Wild+2016}. Traditionally a cut on emission lines has been used to ensure no ongoing star formation, but is now known to bias samples against starbursts that do not decay instantaneously \citep{Tremonti+2007,Wild+2009,Pawlik+2016}, as well as narrow line active galactic nuclei (AGN) which are more prevalent in post-starburst galaxies \citep{Yan+2006,Wild+2007,Pawlik+2018} and galaxies containing shocked gas \citep{Alatalo2016}. 

To avoid these biases Wild et al. \citep{Wild+2007} introduced a method to isolate post-starburst galaxies based on the stellar continuum alone. A Principal Component Analysis (PCA) of the 4000\AA\ region of the optical spectrum identifies three primary principal components that account for 99.5\% of the variation in the shape of spectra, of which we use the first two in this paper. For first component (PC1) measures the shape of the continuum around $4000$\AA\, which is strongly correlated with mean stellar age. PC2 measures the excess Balmer line absorption, meaning that galaxies with large PC2 values have stronger Balmer absorption lines than expected for their mean stellar age. This is typically interpreted as having had a recent starburst, where the enhanced AF/OB star ratio leads to stronger Balmer absorption lines than expected, peaking about 1\,Gyr after the starburst \citep{DresslerGunn1983, Quintero+2004, Nolan+2007, Balogh+2005, Wild+2007, vonderLinden+2010}. The PCA method largely reproduces the method of Kauffmann et al. \citep{Kauffmann+2003} who used the spectral indices D4000 and H$\delta$ to identify galaxies which had recently undergone a burst of star formation. The advantage of PCA over D4000 and H$\delta$ is the increase in signal-to-noise ratio afforded by combining all the higher order Balmer absorption lines, as well as the shape of the continuum, into a single index. It is important to note that this method does not require the starburst to have ceased entirely, unlike the traditional methods that place a hard cut on emission line strength. It is therefore highly effective at identifying an inclusive set of "post-starburst" galaxies, including those that are still forming stars. This allows us to study more generally the causes and impact of starbursts on galaxies, rather than focusing on the more specific class of entirely quenched post-starburst galaxies.  However, it is also worth nothing the disadvantages of the PCA method compared to traditional methods cutting on emission line strength: firstly, the PCA method only cleanly identifies slightly older PSBs ($\gtrsim600$\,Myr); and secondly, dusty star-forming galaxies can contaminate the samples \citep{Pawlik+2018}. With good quality spectra the latter can be identified and removed via their Balmer decrement.

Figure 4 (in the Supplementary Information) shows the distribution of the first two spectral indices for galaxies in the $z=0.1$ snapshot in EAGLE, compared with those measured for galaxies selected from the Sloan Digital Sky Survey \citep[SDSS DR7,][]{Abazajian+2009} with $0.03<z<0.06$, Petrosian $r$-band magnitude less than 17.7, above the same stellar-mass threshold and with signal-to-noise ratio per pixel in the $g$-band, SNR$_g>5$ \citep{Wild+2007}. The median SNR$_g$ of this sample is 14.8. Emission lines are masked during the projection of the SDSS spectra onto the eigenspectra if present. Quiescent galaxies (strong 4000\AA\,break) are found at high values of PC1, while star-forming galaxies form a locus of points at lower PC1 values. Ongoing starbursts occupy the bottom-left corner of the PC1-PC2 distribution as the O and B stars that dominate their spectra have weak Balmer absorption lines; as shown by Wild et al. \citep{Wild+2007} post-starburst galaxies can be selected from the `bump' of the PC1-PC2 distribution at high values of PC2. In Pawlik et al. \citep{Pawlik+2018} we show how this selection compares to the more traditional cuts on the H$\delta$ absorption line and H$\alpha$ emission line equivalent widths. Specifically, using the criteria of \citep{Goto2007} only 9 post-starburst galaxies are found in the SDSS sample presented in Figure 4, compared to 69 using the PCA selection (see below for further details). The PCA-selected PSBs typically have emission line equivalent widths that lie below the star-forming ``main sequence'', but are not (yet) completely absent. 

The EAGLE galaxies show the expected bimodal behaviour in the PC1-PC2 space, in agreement with observations, but there are some differences in the shape of the distribution with respect to SDSS DR7 galaxies. The agreement in PC1 indicates that the distribution of light-weighted mean stellar ages of galaxies in EAGLE is approximately correct, as shown by  \citep{Trayford+2015}. The apparent lack of galaxies in the post-starburst region is simply due to the smaller volume, and we address this point below. The shift towards higher values of PC2 for the EAGLE star-forming galaxies, as well as the reduced width of the PC2 distribution, is caused by several effects. Firstly, we use the same 2-component dust screen model for all EAGLE galaxies, which does not allow for the variation of dust content/geometry seen in real galaxies \citep{Wild+2011,Chevallard+2013,Trayford+2017}. Secondly, the EAGLE spectra are noiseless. Adding Gaussian or white noise to the spectra does not affect the PC amplitudes, while correlated errors and systematic errors that undoubtedly exist in the real spectra are much more difficult to reproduce and so are not added to the mock spectra. Thirdly, while fits to the spectra using spectral fitting package STARLIGHT \citep{2005MNRAS.358..363C} reproduce the PC2 values well, fits using parameterised (i.e. smoothed) star formation and metallicity histories consistently produce PC2 values that are too high. This suggests that the offset is not a fundamental limitation of the spectral synthesis models, but rather a limitation of the star formation and metallicity histories. As shown in von der Linden et al. \citep{vonderLinden+2010}, this can be solved by introducing  fluctuations on timescales of $10^8$ years to the model star formation history, 
possibly suggesting that the EAGLE galaxy disks are not experiencing the level of short-term fluctuations that are observed in galaxies the real Universe. Finally, we find that the SDSS PC2 locus shifts upwards with redshift as a larger fraction of the galaxy light falls within  the fixed 3'' fibre apertures. This leads to spectra more dominated by light from galaxy bulges than their disks and is a well-known effect in all fibre-based surveys, known as ``aperture bias''. Unfortunately it is not possible to correct for aperture bias in these particular measurements.

\subsubsection*{Identifying post-starburst galaxies in EAGLE}

\begin{figure}
\centering
    \includegraphics[width=\columnwidth]{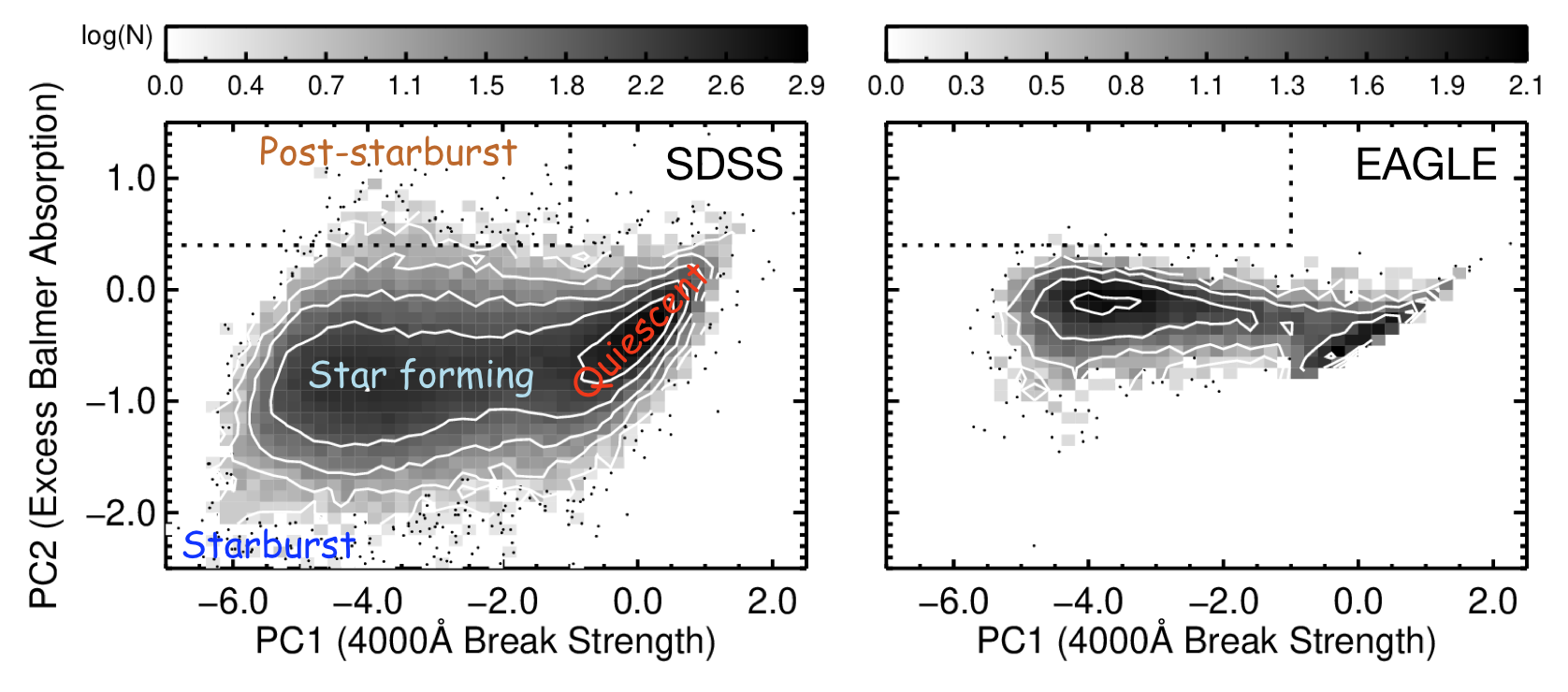}
\caption{ \footnotesize{The distribution of the spectral indices, PC1 and PC2, measured for SDSS galaxies with stellar masses M$_{\star}>10^{9.5}$M$\odot$ and $0.03<z<0.06$ (left) and mock EAGLE galaxies with stellar masses M$_{\star}>10^{9.5}$M$\odot$ in the $z=0.1$ snapshot (right). The greyscale shows the number density of data points (as indicated by the colour bar); individual points are plotted in under-dense regions.}}
\label{fig:PCA}
\vspace{-10pt}
\end{figure}

We place a cut at PC2-$\Delta_{\rm PC2}>0.4$ and PC1$<-1$ to identify a clean set of post-starburst galaxies, where $\Delta_{\rm PC2}$ is the $1\sigma$ error on PC2, and the limits are indicated by dotted lines in Figures \ref{fig:PCA_evolve} and \ref{fig:PCA}. This PC2 cut lies above the locus predicted by models of smooth star formation histories (exponential, delayed exponential or power-law) with dust attenuation typical for these low redshift galaxies; the additional PC1 cut prevents very old/metal rich quiescent galaxies from accidentally falling into the selection box. 
These cuts identify only those galaxies that have undergone a recent starburst, or rapidly and recently shut down their star formation \citep{Wild+2007,vonderLinden+2010}.  We additionally remove SDSS galaxies with large dust contents and strong emission lines that are likely to be dusty star-forming contaminants to the PCA-selected PSB samples \citep{Pawlik+2018}. The smaller volume of EAGLE compared to the SDSS survey means that 7 post-starburst galaxies with stellar masses M$_{\star}>10^{9.5}$M$_\odot$ were identified in the $z=0.1$ snapshot, compared to 69 in our SDSS comparison sample. Accounting for the differing volumes, and assuming a spectroscopic target selection rate of 0.9 in SDSS \citep{Wild+2009}, as well as accounting for 2\% of galaxies lost due to our SNR cut, leads to number densities of $6.7\times10^{-6}$/Mpc$^3$ in SDSS and $7\pm3\times10^{-6}$/Mpc$^3$ in EAGLE, which is remarkably consistent given the limitations of the modelling and data. We note that this consistency between the EAGLE and real Universe in the number density of PCA and photometrically selected post-starburst galaxies remains over a wide range of redshift \citep{Davis2018}. 

The full star formation histories, output every 1\,Myr, for each EAGLE galaxy allow us to improve on these statistics by identifying many more galaxies that enter the post-starburst region in the recent past. We reconstructed the evolving SEDs and calculated the spectral indices of all galaxies in the $z=0.1$ snapshot of the EAGLE simulation, every 50\,Myr, to identify galaxies that entered the post-starburst selection region in the previous 2.5\,Gyr. We identified 124 galaxies ($\approx 2\%$ of the total sample), of which 27\% transition from blue to red within the 2.5\,Gyr probed, 60\% remain within the blue cloud at $z=0.1$, although typically evolve to have a larger PC1 (i.e. lower sSFR) over the measured time interval, and 2\% are in cyclic evolution within the red sequence; the remaining $\approx10\%$ are galaxies found within the post-starburst phase at $z=0.1$, or starting in the post-starburst phase 2.5\,Gyr in the past. We define as ``red'' everything with PC1$>-2$, but additionally inspect the star formation history to ensure galaxies are correctly classified. From visual inspection of the star formation histories, about 50\% of the time the post-starburst phase is subsequent to an evident short-lived (few 100\,Myr) starburst episode; in the other cases a simple truncation of star formation is sufficient to cause the enhanced Balmer absorption lines, often following an extended ($>1$\,Gyr) period of enhanced star formation. 

For this work we selected four galaxies which exemplified each of the different pathways through the post-starburst phase found in the EAGLE simulation (ID: 17473042, 19955965, 9052812, 8154304). In Figure \ref{fig:PCA_evolve} we plot the evolution of their PC spectral indices over the previous 2.5\,Gyr since $z=0.1$. In future work we will compare the dominant parameters leading to the different pathways \citep{DiMatteo+2008,Lagos+2017}, as well as the contribution of post-starburst galaxies to the global build up of the red sequence in comparison to more gentle quenching mechanisms \citep{Rowlands+2018,Trayford+2016}. 

\subsubsection*{Image synthesis and analysis}

The images used in this work were generated using the SKIRT Monte Carlo radiative transfer code \citep{Baes+2003,CampsBaes2015}, with GALAXEV \citep{BruzualCharlot2003} and MAPPINGS III \citep{Groves+2008} spectra, as described in \citep{Trayford+2017}. In contrast to Trayford et al. \citep{Trayford+2017}, we do not perform the resampling of the recent star formation, as this smooths the star formation histories over 100\,Myr which impacts the spectral features of the post-starburst galaxies. Images with 0.396''/pixel sampling were created, both with and without the effect of attenuation of light due to dust. From this we find that dust has no significant effect on the morphology parameters measured here (see Figure \ref{fig:evolution}). 
The images were then made to mimic the typical SDSS $r$-band images by adding effects of noise and point spread function smoothing (see \citep{Pawlik+2018} for more details). We chose to create all mock images at a fixed redshift of $z=0.04$, in order to focus on the physical evolution alone, independent of image redshifting effects. 

To characterise the morphological evolution of EAGLE galaxies through the post-starburst phase we analysed their mock images using 1) the shape asymmetry parameter, $A_{S}$, which acts as an indicator of ongoing/past mergers when its value becomes larger than 0.2 \citep{Pawlik+2016}; 2) the concentration index, $C$, which distinguishes between galaxies with dominant disky ($C\approx3$) and spherical ($C>4$) components \citep{Bershady+2000}, but may also increase temporarily during a centralised starburst. Both $A_{S}$ and $C$ were measured using the code described in \citep{Pawlik+2016}.

\section*{Data availability}
The raw data used in this paper are available to download from the EAGLE public database \url{http://icc.dur.ac.uk/Eagle/database.php}. The case study galaxies presented here can be identified via their unique {\sc galaxyid}. The database is described in \citep{McAlpine+2016} and \citep{EagleDATA}. The eigenvectors and PC amplitudes for SDSS DR7 spectra presented in Figures 3 and 4 are available for download from \url{http://www-star.st-and.ac.uk/~vw8/downloads/DR7PCA.html}. The PCA code and morphological analysis code are available for download from \url{https://github.com/SEDMORPH}. The spectral synthesis models used to build the spectra from the EAGLE star formation histories are available for download from \url{http://www.bruzual.org/bc03/Updated_version_2016/}.

\end{document}